\documentclass[trackchanges]{aastex701}
\usepackage{amsmath,amssymb} 
\begin{document}

\title{The rise of the black hole X-ray binary AT\,2019wey observed with TESS}

\author[orcid=0009-0006-3909-7411,sname='Alyana Jusino']{Alyana Jusino}
\affiliation{Department of Astrophysics, American Museum of Natural History, Central Park West at 79th Street, New York, NY 10024, USA}
\affiliation{CUNY City College of the City University of New York, New York, New York 10031, USA}
\email[show]{alyanajusino@gmail.com}  

\author[orcid=0000-0002-8989-0542, sname='Kishalay De']{Kishalay De} 
\affiliation{Department of Astronomy and Columbia Astrophysics Laboratory, Columbia University, 550 W 120th St. MC 5246, New York, NY 10027, USA}
\affiliation{Center for Computational Astrophysics, Flatiron Institute, 162 5th Ave., New York, NY 10010, USA}
\email[show]{kd3038@columbia.edu}

\author[orcid=0000-0003-3062-4773, sname='Andrea Antoni']{Andrea Antoni}
\affiliation{Center for Computational Astrophysics, Flatiron Institute, 162 5th Ave., New York, NY 10010, USA}
\email[show]{aantoni@flatironinstitute.org}

\begin{abstract}
Black hole X-ray binaries (BHXRBs) have traditionally been discovered by X-ray surveys with cadences of hours to days. However, large optical time-domain surveys now provide novel avenues for early detection and insights into their elusive outburst triggering mechanisms. We present early-time light curves of the BHXRB AT 2019wey serendipitously observed by the Transiting Exoplanet Survey Satellite (TESS). The TESS images are sampled at 30 minute cadence from $\approx2$\,d prior to $\approx25$\,d after outburst, providing the highest time resolution optical rising phase observations of any known BHXRB. We fit a piece-wise power law to the rising light curve, finding an outburst onset time of MJD $58817.86\pm0.09$ and power-law rise index $n=0.74\pm0.04$. The onset time precedes all ground-based optical detections, and suggests that the optical rise began after the start of the faint X-ray brightening in MAXI data. We search for periodic high frequency modulation and detect none exceeding amplitude $\approx0.48$\,mJy at periods of $\gtrsim1$\,h at 90\% confidence.
\end{abstract}



\section{INTRODUCTION}
\label{sec:intro}

Low-mass X-ray Binaries (LMXBs) are binary star systems that contain a compact object, either a neutron star (NS) or a black hole (BH), with a low-mass companion donor star. Gas from the donor is transferred through an accretion disk onto the compact object, producing X-ray emission \citep{Bahramian_2023, refId0,2018MNRAS.480....2T}. LMXBs are commonly transient systems, with long quiescent intervals interrupted by dramatic outbursts -- when the accretion rate onto the compact object suddenly rises, causing the source to brighten across the electromagnetic spectrum \citep{1998ApJ...493..351K}. During these outbursting phases, the source luminosity can increase by several orders of magnitude within a few days, and then undergo exponential decay over the course of few weeks to months, before returning to quiescence \citep{refId0}. A key insight into understanding LMXBs and their accretion mechanism lies in the initial outburst phase, which is often missed due to its rapid onset \citep{Russell_2019} -- but can help reveal the poorly understood triggers of the outbursts.

AT 2019wey is a recently discovered BH LMXB candidate, first identified in December 2019 by the ATLAS optical survey \citep{2019TNSTR2553....1T} and later classified via its X-ray emission detected by the Spektrum-Roentgen-Gamma (SRG) mission \citep{Yao_2020, Mereminskiy2022}. The source was observed by NASA’s Transiting Exoplanet Survey Satellite (TESS) as it entered outburst, capturing the rise with high photometric precision and nearly uninterrupted 27-day coverage \citep{2015JATIS...1a4003R}. Here, we present the TESS light curve of the early outburst and results from preliminary analysis of the light curve.

\bigskip
\section{OBSERVATIONS AND DATA ANALYSIS} \label{sec:style}

AT 2019wey was observed by TESS in Sector 19, with full-frame images obtained using a 30-minute cadence. These observations captured the initial stages of the outburst, with coverage beginning on MJD 58815.598816 -- approximately $9$\,d prior to the first report on the Transient Name Server \citep{2019TNSTR2553....1T}. TESS light curves of AT 2019wey were extracted using the open-source Python package TESSreduce, which performs automated background subtraction, photometric calibration, image alignment, and subtraction against the pre-outburst template \citep{2021arXiv211115006R}. We also binned the data into six-hour intervals to probe longer-timescale trends. The 30-minute cadence data was modeled using a piecewise function that consists of a flat pre-outburst baseline and a power-law rise: 

\begin{equation}
\\F(t)=
    \begin{cases}
        C \text{ ,} & t < t_0\\
        C+A\cdot(t-t_0)^n  \text{, } & t \geq t_0
    \end{cases}
\end{equation}

The model fit was then performed with the curve fit function in scipy to derive the best-fit parameters and uncertainties from the covariance matrix. After obtaining a best-fit model, we calculated the residual light curve by subtracting the modeled flux from the observed flux. By applying the Lomb-Scargle periodogram on the residuals, we searched for low-level periodic modulation imprinted on the long-term light curve. 

\begin{figure*}[ht!]
\plotone{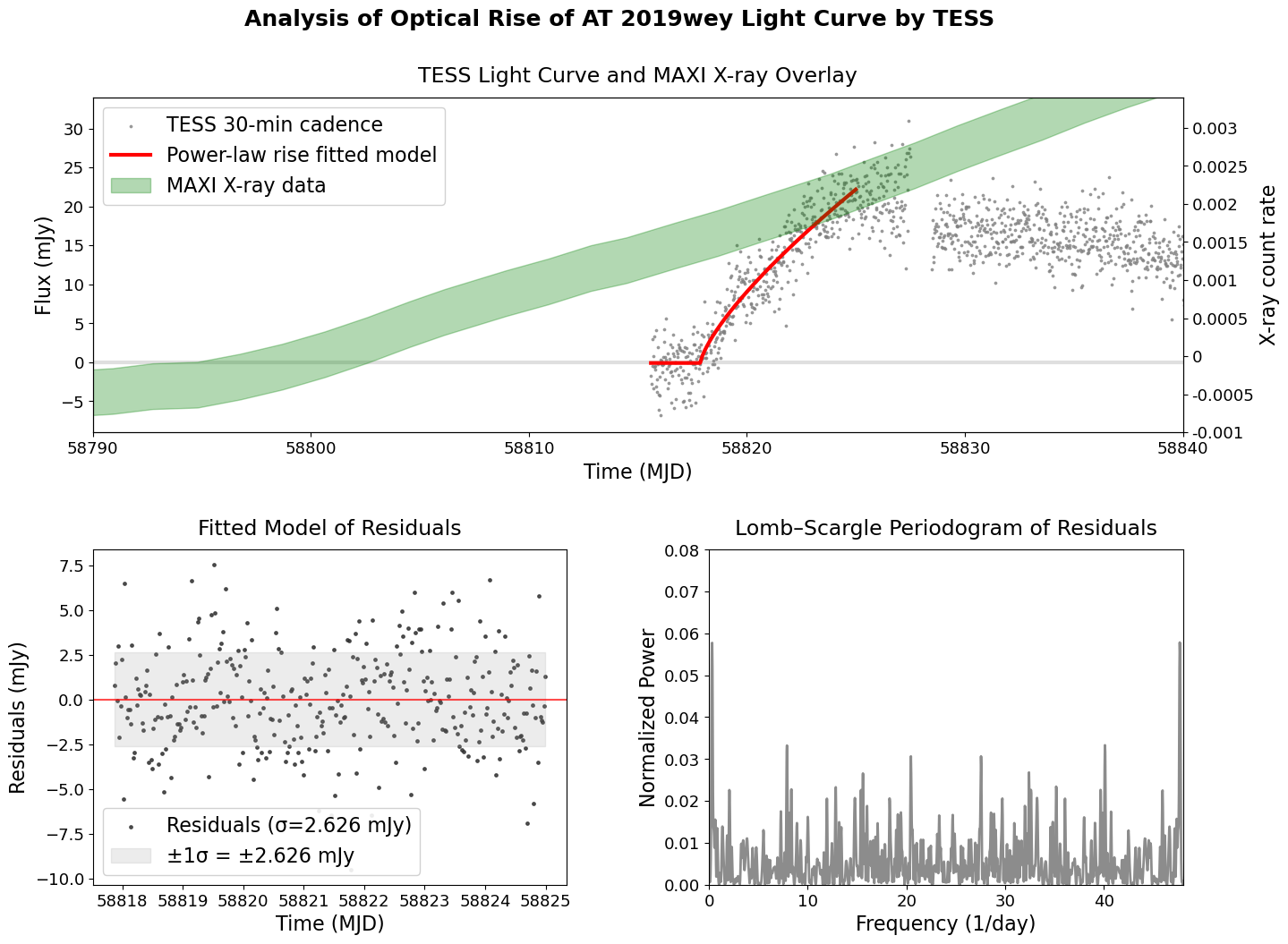}
\caption{Optical rise of AT\,2019wey observed by TESS overlaid with \citet{Yao_2020} model. ({\it Top panel}) Reduced TESS light curve of the AT\,2019wey outburst observed in Sector 19. Gray dots are individual measurements from difference imaging on the TESS full-frame images, with the red being the power-law rise model. The red line is the best-fit piecewise power-law function fit to the rising light curve. The green shaded region indicates the best-fit MAXI X-ray evolution found by \citet{Yao_2021}. ({\it Bottom left panel}) Residuals of the TESS data after subtracting the best-fit model (red dashed line centered at zero) for the light curve rise. The gray shaded region shows the $\pm 1\sigma$ standard deviation of the residuals. ({\it Bottom right}) Lomb-Scargle periodogram of the residuals shown in the bottom left panel. }
\label{fig:1}
\end{figure*}

\bigskip
\section{RESULTS AND CONCLUSION} \label{sec:floats}

We obtained the best-fit parameters, along with their 1$\sigma$ uncertainties of $A = 5.22 \pm 0.39$, $t_0 = 58817.86 \pm 0.09$, $n = 0.74 \pm 0.04$, and $ C = -0.10 \pm 0.26$ (consistent with zero). The best-fitting model is shown in Figure \ref{fig:1}. The TESS cadence constrains the onset time of the outburst to within approximately 2 hours. The derived onset time precedes the first detection of the source by the ZTF (which first detected the source on MJD 58819.22) and ATLAS surveys (which reported a non-detection on MJD 58818.51). Comparison with the MAXI X-ray light curve (Figure \ref{fig:1}; from \citealt{Yao_2021}) confirms that the X-ray brightening also began prior to the first detection from ZTF and TESS (see Figure \ref{fig:1}), although the low signal-to-noise ratio of the MAXI data precludes a quantitative comparison. Nominally, earlier X-ray emission than the optical onset supports an inside-out outburst scenario, in which thermal instability begins in the inner accretion disk and then moves outward \citep{2018MNRAS.480....2T, Russell_2019}. 

Previously \citet{Yao_2021} reported a search for periodic modulation using ZTF data and follow-up fast imaging using the Palomar 200-inch telescope. A possible 1.3-hour period was identified on the fading phase of the light curve, but excluded because it was not observed to repeat. We use the TESS high cadence data to search for low-amplitude periodic modulation superimposed on the rising phase. We subtract out the best-fit model for the rising light curve, and use the residuals to search for periodic modulations using a Lomb-Scargle periodogram (see Figure \ref{fig:1}). No significant periodic modulations were detected exceeding an amplitude of 0.48 mJy at 90\% confidence in the sampled frequency range (corresponding to periods of $\approx 1$\,h to $\approx 14$\,d). 

\bigskip
\bigskip
\section{ACKNOWLEDGMENTS}
We acknowledge support from AstroCom NYC, the Simons Foundation through award \#00533845 and National Science Foundation Partnerships in Astronomy and Astrophysics Research and Education through grant AST-2219090.


\bibliography{references}
\bibliographystyle{aasjournalv7}



\end{document}